\documentclass[aps,prd,twocolumn,10pt,superscriptaddress,nofootinbib,nobibnotes,longbibliography]{revtex4-1}
\usepackage{amsmath}
\usepackage{mathrsfs}
\usepackage{amssymb}
\usepackage{mathtools}
\usepackage{amsmath}
\usepackage{csquotes}
\usepackage{booktabs}
\usepackage{lipsum} %
\usepackage{adjustbox}
\PassOptionsToPackage{colorlinks=true,linkcolor=blue,citecolor=red,urlcolor=magenta}{hyperref}
\usepackage{orcidlink}
\RequirePackage{graphicx}
\usepackage{microtype}
\usepackage{graphicx} %
\usepackage{subcaption} %
\usepackage{stfloats}
\usepackage{placeins}
\usepackage{graphicx}
\usepackage{dcolumn}
\usepackage{bm}

\usepackage{xcolor}

\begin{document}

\title{Hamiltonian formulation of Carrollian Maxwell theory in Deformed Light‑cone Kaluza–Klein-like Null reduction}

\author{Limin Zeng\orcidlink{0009-0003-5703-2687}}
\email{zenglimin25@mails.ucas.ac.cn}
\affiliation{
   School of Fundamental Physics and Mathematical Sciences,\\
   Hangzhou Institute for Advanced Study, UCAS, Hangzhou 310024, China
 }
 \affiliation{Institute of Theoretical Physics, Chinese Academy of Sciences, Beijing 100190, China}
 \affiliation{University of Chinese Academy of Sciences, Beijing 100049, China}
 
\date{\today}
\begin{abstract}
We construct magnetic and electric Carrollian Maxwell theories by performing Kaluza–Klein-like null reduction of a complex Maxwell field in a Bargmann deformed light-cone background with manifest gauge symmetry. {The procedure preserves the original first‑class U(1) Gauss constraint in the standard  Carrollian Maxwell theories and deformed Carrollian Maxwell theories with a decoupled electric/magnetic scalar $A_-$. We give the scaling of the parent gauge parameter and show that the $U(1)$ symmetry is inherited from the parent theory and not introduced after the limit. We emphasize that $A_-$ can be rendered non-dynamical by imposing additional ad hoc constraints. Since $A_-$ is invariant under the original $U(1)$ symmetry, removing it is a constraint and not a gauge choice, but the resulting theory is the same as the one reached by imposing the light-cone gauge $A_- = 0$ prior to the KK-like null reduction. For an electric pseudo-coupled theory without gauge symmetry, after imposing the ad hoc constraints the theory becomes the standard electric Maxwell theory together with a decoupled zero-energy scalar $A_-$.} Furthermore, our work provides an explicit example of the correct application of this approach, thereby broadening the scope of its applicability to gauge theories. 
\end{abstract}

\maketitle

\section{Introduction}
\label{sec:1}

Carrollian physics \cite{Levy-Leblond:1965dsc,SenGupta:1966qer,deBoer:2023fnj,Hansen:2021fxi,Ciambelli:2019lap,Bagchi:2025vri} has attracted considerable attention in recent years due to flat holography\cite{Bagchi:2022emh,Donnay:2022aba,Costello:2022jpg,Ruzziconi:2026bix}. 
The systematic construction of Carrollian field theories\cite{Gupta:2020dtl,Bagchi:2022eav,Koutrolikos:2023evq,Sharma:2025rug} from relativistic parents is a central problem, and several complementary strategies have been developed. The first method, group contraction\cite{Henneaux:2021yzg}, {starts from the Hamiltonian formulation of Lorentz‑invariant theories and introduces independent $c\to0$ scalings for the canonical variables while preserving the symplectic structure, thereby obtaining electric and magnetic Carroll contractions from the same Lorentzian parent theory.} A second approach is Bargmann‑based null reduction \cite{Chen:2023pqf}\cite{Majumdar:2025juk}. Ref. \cite{Chen:2023pqf} constructs two distinct Bargmann‑invariant parent actions: one with the null vector $\xi^\beta\xi^\delta$ contracted into the field strength and the other with the Bargmann metric $G^{\beta\delta}$, whose null reductions yield the electric and magnetic sectors respectively. {However, Ref. \cite{Majumdar:2025juk} starts from a light‑cone action in Minkowski spacetime, adds a deformation term to controllably reduce the Poincaré invariance to a Bargmann subgroup with a tunable deformation parameter $\alpha$, and then projects the theory onto a null hypersurface by means of a smearing function, thus obtaining a Carrollian theory in one lower dimension. The imposition of the light‑cone gauge condition $A_-=0$ on the gauge field, both before and after the reduction, ensures that the resulting theory is precisely the standard Carrollian Maxwell theory.} A third avenue\cite{Lenz:1991sa,Julia:1994bs,Saha:2025wkr,Zeng:2026zcj}, starts from an action in deformed light‑cone coordinates (Bargmann background), and compactifies the null direction by expanding the fields in Kaluza–Klein-like(KK-like) modes\cite{Salam:1981xd}, and finally takes $c\to0$ explicitly to obtain Carroll theory. The original idea of considering Carroll symmetry lightlike reduction has been proposed in \cite{Duval:1990hj} and have been further developed in \cite{Duval:2014uoa}.

The third avenue shares the same geometric background with \cite{Majumdar:2025juk}: the deformed light-cone metric is equivalent to a Bargmann coordinate choice on standard Minkowski spacetime. The difference is that \cite{Majumdar:2025juk} treats $\alpha$ as a tunable deformation parameter that breaks Poincaré invariance down to Bargmann and acts as a selector between the electric and magnetic sectors, whereas we treat it as a fixed parameter to serve solely to eliminate the second-class constraints induced by the standard light-cone coordinates. Our approach is equivalent to adding a deformation term to the light-cone Lorentzian action, thereby yielding the Bargmann theory. However, we separate the different magnetic and electric Carroll sectors by means of independent scalings of canonical variables in Hamiltonian phase space. {By not imposing the light‑cone gauge condition, we obtain a family of non‑standard Carrollian theories, in which the gauge field can coexist with an extra scalar field $A_-$ in a decoupled manner.}

When applied to a complex Maxwell field in a manifestly gauge‑invariant formulation, the third scheme has been shown to produce a theory of free complex scalars without gauge symmetry \cite{Zeng:2026zcj,zeng2026gauge}. However, we emphasize that both \cite{Zeng:2026zcj} and \cite{zeng2026gauge} are formulated within the Lagrangian framework and do not rely on any additional $c$-rescaling to yield an free scalar theory. Under these circumstances, the degeneration of the gauge symmetry of the parent theory is accurately addressed in Section 3.4 of Ref. \cite{zeng2026gauge}.

In this paper we demonstrate that the degeneration of gauge symmetry is not an intrinsic feature of KK-like null reduction, but a consequence of the unified scaling assumption. By passing to the Hamiltonian formulation\cite{Dirac:1950pj,Henneaux:1992ig} before taking the $c\to0$ limit and allowing the canonical momenta to scale independently of their conjugate coordinates we are able to preserve an {original} first‑class U(1) Gauss constraint in the standard Carrollian Maxwell theories and deformed Carrollian Maxwell theories with a decoupled electric/magnetic scalar $A_-$. We verify this by giving the scaling of the parent gauge parameter, $\tilde\epsilon = c^{\rho}\epsilon'$ with $\rho = \gamma_A$, and by checking that the parent transformations stay finite in the limit. 
{Even for an electric pseudo-coupled theory without gauge symmetry obtained from our particular choice of scalings,  imposing an ad hoc Gauss constraint reduces the theory to electric Maxwell theory with a decoupled zero-energy scalar.} These resulting theories retain a non‑trivial gauge structure and are legitimate Carrollian Maxwell theories. Prior to this work, this method had not yet yielded a consistent Carrollian gauge theory.

The paper is organised as follows. Section \ref{sec:2} reviews the parent theory and sets up the Hamiltonian null reduction without fixing the light‑cone gauge. In Section \ref{sec:3}, we take the Carrollian limit with independent canonical scalings, derive different magnetic and electric canonical actions, and analyze the gauge symmetry and physical degrees of freedom. Section \ref{sec:4} presents the correct procedure for obtaining Carrollian theories via the deformed light-cone KK-like null reduction method, and clarifies misconceptions in previous studies. {Besides, we clearly distinguish the features of our approach compared to the previous method employed in Ref. \cite{Majumdar:2025juk}. Section \ref{sec:5} summarizes the whole work and provides an outlook on future research directions.}

\section{Parent Theory and Deformed KK-like Light-Cone Null Reduction}
\label{sec:2}

We consider a complex Maxwell field $A_\mu$ in $(d+1)$ spacetime dimensions. The metric is the deformed light‑cone metric \cite{Lenz:1991sa}
\begin{equation}
\begin{aligned}
ds^2 &= -2dx^+dx^- + (dx^-)^2 + \delta_{ij}dx^i dx^j, \\
g^{\mu\nu} &= \begin{pmatrix}
-1 & -1 & 0 \\
-1 &  0 & 0 \\
 0 &  0 & \delta^{ij}
\end{pmatrix}.
\end{aligned}
\tag{2.1} \label{eq:2.1}
\end{equation}
with $i,j=1,\dots,d-1$. The action is
\begin{equation}
S = -\frac14 \int dx^+ dx^- d^{d-1}x\; F_{\mu\nu}^* F^{\mu\nu},\tag{2.2} \label{eq:2.2}
\end{equation}
where $F_{\mu\nu}= \partial_\mu A_\nu - \partial_\nu A_\mu$. Invariance under the U(1) gauge transformation $\delta A_\mu = \partial_\mu\epsilon$, $\delta A_\mu^* = \partial_\mu\epsilon^*$ is manifest.

We introduce the KK-like mode expansion 
\begin{equation}\begin{aligned}A_\mu &= \int dp_{-}e^{-ip_-x^-}\tilde A_\mu(x^+,x^i).
\end{aligned}\tag{2.3} \label{eq:2.3}
\end{equation}
The $\int dx^-$ yields $2\pi\delta(p_- - p_-')$, eliminating one momentum integral. The remaining $\int dp_- = c\int dm$ together with $dx^+ = c\,d\tau$ yields an overall factor $c^2$. 
Inserting these expansions into the action and discarding the overall $2 \pi \int dm$,  gives the effective action:
\begin{equation}
\begin{split}
S =& c^2\int d\tau d^{d-1}x \mathcal{L},\\
\mathcal{L} = &\frac{1}{2}\big| imc\tilde{A}_+ + \partial_+\tilde{A}_- \big|^2 + \frac{1}{2}\sum_i \big| \partial_+\tilde{A}_i - \partial_i\tilde{A}_+ \big|^2 \\
&+ \sum_i \operatorname{Re}\!\big[ (\partial_i\tilde{A}_- + imc\tilde{A}_i)(\partial_i\tilde{A}_+^* - \partial_+\tilde{A}_i^*) \big] \\
&- \;\frac{1}{4}\sum_{i,j} \big| \partial_i\tilde{A}_j - \partial_j\tilde{A}_i \big|^2.
\end{split}
\tag{2.4} \label{eq:2.4}
\end{equation}
Choosing $x^+$ as the evolution parameter, the canonical momenta are
\begin{equation}\begin{aligned}
\tilde\Pi^\mu &\equiv \frac{\partial\mathcal{L}}{\partial(\partial_+ \tilde A_\mu)},\\
\tilde\Pi^+ &\approx 0, \\
\tilde\Pi^- &= \frac{1}{2}\bigl(\partial_+ \tilde A_-^* - imc \tilde A_+^*\bigr), \\
\tilde\Pi^i &= \frac{1}{2}\bigl(-\partial_i \tilde A_+^* + \partial_+ \tilde A_i^* - \partial_i \tilde A_-^* + imc \tilde A_i^*\bigr),
\end{aligned}
\tag{2.5} \label{eq:2.5}
\end{equation}
and their conjugates.
We can obtain the time derivative of the fields:
\begin{equation}
\begin{aligned}
\partial_+ \tilde A_- &= 2\tilde\Pi^{-*} - imc \tilde A_+ ,\\
\partial_+ \tilde A_i &=2\tilde \Pi^{i*} + \partial_i \tilde A_+ + \partial_i \tilde A_- + i m c\tilde A_i .
\end{aligned}
\tag{2.6} \label{eq:2.6}
\end{equation}

The canonical Hamiltonian density takes the form
\begin{equation}\begin{aligned}
\mathcal{H}_c =& \tilde\Pi^\mu \partial_+ \tilde A_\mu + \tilde\Pi^{*\mu} \partial_+ \tilde A_\mu^* - \mathcal{L}\\=&2|\tilde\Pi^-|^2 + \frac12 \sum_i \big| 2\tilde \Pi^i + \partial_i \tilde A_-^* - i m c\tilde A_i^* \big|^2 \\&+ \frac14 \sum_{i,j} |\tilde F_{ij}|^2 - \tilde A_+ \mathcal{G} - \tilde A_+^* \mathcal{G}^* .
\end{aligned}
\tag{2.7}\label{eq:2.7}
\end{equation}
We have dropped some surface terms and define
\begin{equation}
\mathcal{G} \equiv \partial_i \tilde\Pi^i + i m c \tilde\Pi^-
\tag{2.8}\label{eq:2.8}
\end{equation}
The total Hamiltonian reads 
\begin{equation}\begin{aligned}
H_T = \int dx^- d^{d-1}x \Big( \mathcal{H}_{c} + u \, \tilde \Pi^+ + u^* \, \tilde \Pi^{+*} \Big).
\end{aligned}\tag{2.9} \label{eq:2.9}\end{equation}
where $u$ is the multiplier.
One can check the secondary constraint (and its conjugate) comes from: 
\begin{equation}\begin{aligned}
\partial_+ \tilde\Pi^+(x) = \{\tilde\Pi^+(x), H_T\}=- \frac{\delta H_T}{\delta \tilde A_+(x)} =\mathcal{G}(x) \approx 0,
\end{aligned}\tag{2.10} \label{eq:2.10}\end{equation}
and $\tilde A_+$ naturally becomes the multiplier for $\mathcal{G}$. There are no more constraints since \begin{equation}\begin{aligned}\{\mathcal{G}(x), H_T\} \approx 0.
\end{aligned}\tag{2.11} \label{eq:2.11}\end{equation}
These constraints are all first-class constraints. Therefore, the physical phase space has complex dimension $2(d-1)$ per space point, corresponding to the $d-1$ transverse photon polarisations.

And the canonical action is
\begin{equation}\begin{aligned}
S_H = c^2\int d\tau d\vec{x} \, \Big[2c^{-1}\operatorname{Re}( 
\tilde\Pi^\mu \partial_\tau \tilde A_\mu)  - \mathcal{H}_c -2\operatorname{Re}( u \tilde\Pi^+) 
\Big] .
\end{aligned}\tag{2.12} \label{eq:2.12}\end{equation}
It can be verified that the action is invariant under the following gauge transformations:
\begin{equation}\begin{aligned}
\delta \tilde{A}_+ &= \partial_+ \tilde{\epsilon} , 
\; \delta \tilde{A}_- = -i m c \tilde{\epsilon} , \\
\delta \tilde{A}_i &= \partial_i \tilde{\epsilon} , \;\;\
\delta \tilde{\Pi}^\mu = 0 , \\
\delta u &= \partial_+^2 \tilde{\epsilon} .
\end{aligned}\tag{2.13} \label{eq:2.13}\end{equation}
It is generated by the following first-class constraints acting as generators (understood as also {null} reducing $\tilde\epsilon$, and in the sense of acting on the original canonical action):
\begin{equation}\begin{aligned}
G[\tilde{\epsilon}] =& \int d\vec{x} \, \Bigl[ -\tilde{\epsilon}(x)\,\mathcal{G}(x) + (\partial_+ \tilde{\epsilon}(x))\,\tilde{\Pi}^+(x) \Bigr] + \text{c.c.},\\&
\{\tilde{A}_\mu(x), G\} = \delta \tilde{A}_\mu(x).
\end{aligned}\tag{2.14} \label{eq:2.14}\end{equation}

We do not impose any gauge condition on $\tilde A_-$ now; it remains a dynamical canonical variable together with its conjugate momentum $\tilde\Pi^-$. {This is precisely why a gauge‑invariant scalar field $A_-$ could still appear in the 
nonstandard magnetic or electric Carroll theory later. And to highlight the Carrollian features in what follows, we denote $A_+$ as $A_\tau$ and $\Pi^+$ as $\Pi^\tau$.}

\section{Carrollian limit with independent canonical scalings}
\label{sec:3}

Throughout this section and the appendices we assume that the fields fall off sufficiently fast at spatial infinity. This condition removes the harmonic modes that would otherwise survive some of the additional constraints introduced below, and it is what makes the degree-of-freedom counts well defined.

We also define $H$ as the canonical Hamiltonian of the theory under discussion. It is the spatial integral of the reduced $\mathcal{H}_c$ of Eq. \eqref{eq:2.7}. Apart from the kinetic terms, the only contributions it leaves out are those carrying the multipliers $u$ and $w$, such as $-u\Pi^\tau$ and $-w\Pi^\sigma$. For Eq. \eqref{eq:3.4} this gives $H=\int d\vec{x}\,[\frac14F_{ij}F^{ij}-A_\tau\partial_i\Pi^i]$. Whenever a bracket with $H$ is used below we write $H$ out for that theory. All Poisson brackets are taken with this $H$.

\subsection{General scaling ansatz}

We allow independent scalings for the canonical pairs and the mass parameter:
\begin{equation}\begin{aligned}
\tilde A_i      &= c^{\gamma_A} A'_i , \qquad & \tilde\Pi^i    &= c^{\gamma_\Pi} \Pi'^{\,i},\\\tilde
A_-      &= c^{\beta_A}  A'_- , \qquad &\tilde \Pi^-    &= c^{\beta_\Pi}  \Pi'^{\,-},\\ \tilde
A_\tau   &= c^{\alpha_A}  A'_\tau , \quad & \tilde\Pi^\tau    &= c^{\alpha_\Pi}  \Pi'^{\,\tau},\\
u        &= c^{\mu}      u',  \qquad & m   &=c^{\nu}m',\\
\tilde\epsilon &= c^{\rho}\epsilon' , & &
\end{aligned}\tag{3.1} \label{eq:3.1}\end{equation}
We keep the physical mass fixed, so $m$ is $c^0$ (i.e. we set the scaling exponent $\nu=0$). The primed fields are understood to remain finite as $c\to0$. We also scale the gauge parameter of the parent theory, see the last line of Eq. \eqref{eq:3.1}. The exponent $\rho$ is fixed at the end of this subsection. 
Substituting \eqref{eq:3.1} into $S_H$ \eqref{eq:2.12} and collecting powers of $c$ yields the exponents listed in Table \ref{tab:1}. {In the table we omit the prime notation on the variables in Eq. \eqref{eq:3.1}.}
\begin{table}[h]
  \centering
  \renewcommand{\arraystretch}{1.3}
  \begin{tabular}{lc}
    \toprule
    \textbf{Term} & \textbf{Scaling exponent} \\
    \midrule
    \multicolumn{2}{l}{\textit{Kinetic terms (from $2\operatorname{Re}(\Pi^\mu \partial_+ A_\mu)$)}} \\
    $\Pi^i \partial_\tau A_i$       & $1 + \gamma_\Pi + \gamma_A$ \\
    $\Pi^- \partial_\tau A_-$       & $1 + \beta_\Pi + \beta_A$ \\
    $\Pi^\tau \partial_\tau A_\tau$    & $1 + \alpha_\Pi + \alpha_A$ \\
    \midrule
    \multicolumn{2}{l}{\textit{Potential and gradient terms (from $\mathcal{H}_c$)}} \\
    $|\Pi^-|^2$                     & $2 + 2\beta_\Pi$ \\
    $|\Pi^i|^2$                     & $2 + 2\gamma_\Pi$ \\
    $\operatorname{Re}(\Pi^i \partial_i A_-)$ & $2 + \gamma_\Pi + \beta_A$ \\
    $|\partial_i A_-|^2$            & $2 + 2\beta_A$ \\
    $|F_{ij}|^2$                    & $2 + 2\gamma_A$ \\
    \midrule
    \multicolumn{2}{l}{\textit{Mass-induced terms (from explicit $mc$ in $\mathcal{H}_c$)}} \\
    $m^2 c^2 |A_i|^2$              & $4 + 2\gamma_A$ \\
    $\operatorname{Im}(\Pi^i\, m c\, A_i^*)$ & $3 + \gamma_\Pi + \gamma_A$ \\
    $\operatorname{Re}(\partial_i A_-\, m c\, A_i^*)$ & $3 + \beta_A + \gamma_A$ \\
    $A_\tau\, m c\, \Pi^-$         & $3 + \alpha_A + \beta_\Pi$ \\
    \midrule
    \multicolumn{2}{l}{\textit{Constraint-multiplier terms}} \\
    $A_\tau \partial_i \Pi^i$      & $2 + \alpha_A + \gamma_\Pi$ \\
    $u \Pi^\tau$                      & $2 + \mu + \alpha_\Pi$ \\
    \bottomrule
  \end{tabular}
  \caption{Scaling exponents of each term in $S_H$, Eq. \eqref{eq:2.12}.}
  \label{tab:1}
\end{table}

Because we want a well‑defined action in the limit $c\to0$, all terms we wish to keep must have $c$ exponent $=0$; unwanted terms must have $c$ exponent $>0$.

We now fix the scaling of the gauge parameter. The parent transformations \eqref{eq:2.13} must reduce to those of the reduced theory once the parent parameter carries a definite power of $c$. Combining Eq. \eqref{eq:2.13} with $\partial_+ = c^{-1}\partial_\tau$ and the ansatz \eqref{eq:3.1}, the transformation of $A_i$ reads
$$\delta A'_i = c^{-\gamma_A}\,\partial_i\bigl(c^{\rho}\epsilon'\bigr) = c^{\rho-\gamma_A}\,\partial_i\epsilon' .$$
This is finite and equal to $\partial_i\epsilon'$ only for $\rho = \gamma_A$. The same value keeps the other transformations finite for every scaling used in Section \ref{sec:3}. For $A_\tau$ we get $\delta A'_\tau = c^{\rho-\alpha_A-1}\partial_\tau\epsilon'$, and for the multiplier $\delta u' = c^{\rho-\mu-2}\partial_\tau^2\epsilon'$. For the schemes of Eqs. \eqref{eq:3.2}, \eqref{eq:3.14}, \eqref{eq:3.17}, \eqref{eq:3.26} and \eqref{eq:3.29} both exponents vanish, so the two transformations keep their standard form $\delta A'_\tau = \partial_\tau\epsilon'$ and $\delta u' = \partial_\tau^2\epsilon'$. These are the transformations that go with the constraints of Eqs. \eqref{eq:3.13} and \eqref{eq:3.25}. The remaining schemes discard $A_\tau$ and $u$, so they impose no condition. For the scalar we get $\delta A'_- = -im\,c^{1+\rho-\beta_A}\epsilon'$, which vanishes in the limit whenever the scalar is kept, since the exponent $1+\gamma_A-\beta_A$ is at least $1$ in every such scheme. In the standard magnetic scaling of Eq. \eqref{eq:3.2} the scalar is absent from the reduced action, so its transformation is not required to vanish. The constraints are inherited in the same way. In the schemes where the Gauss term survives, the parent constraint $\mathcal{G}=\partial_i\tilde\Pi^i+imc\tilde\Pi^-$ has its mass term suppressed by at least one extra power of $c$ relative to the field term, so it reduces in the limit to the Gauss constraint $\partial_i\Pi'^i\approx0$ of Eqs. \eqref{eq:3.10} and \eqref{eq:3.25}. This shows that the U(1) symmetry of the reduced theories is inherited from the parent theory, and is not introduced after the limit.

\subsection{Magnetic contraction}

\subsubsection{Standard contraction}
For the magnetic contraction we wish to keep the magnetic energy $\propto|F'_{ij}|^2$ while suppressing the electric energy $\propto|\Pi^{\prime i}|^2$. The kinetic terms must remain finite, and the Gauss‑multiplier term should also survive to enforce the constraint. Examining Table \ref{tab:1}, a consistent choice of exponents is
\begin{equation}\begin{aligned}
  \gamma_A &= -1, & \gamma_\Pi &= 0, & \alpha_A &= -2, \\
  \alpha_\Pi &= 1, & \beta_A &= 0, & \beta_\Pi &= 0, \qquad
  \mu = -3.
\end{aligned}
\tag{3.2} \label{eq:3.2}
\end{equation}

In the limit $c\to0$, the action \eqref{eq:2.12} therefore reduces to
\begin{equation}\begin{aligned}
S_{\rm mag} = \int d\tau\, d\vec{x}\;
\Big[ &2\operatorname{Re}(\Pi'^{\,i}\partial_\tau A'_i)+2\operatorname{Re}(\Pi'^{\,\tau}\partial_\tau A'_\tau)
\\&- \frac{1}{4} \sum_{i,j} |F'_{ij}|^2
+ A'_\tau \partial_i \Pi'^{\,i} \\&+  A'_\tau{}^* \partial_i \Pi'^{\,i*}
- u' \Pi'^{\,\tau} - u'{}^* \Pi'^{\,\tau*} \Big].
\end{aligned}
\tag{3.3} \label{eq:3.3}
\end{equation}
{Restricting the complex theory to a single real copy (neglecting the prime of each variables) yields the real action}
\begin{equation}\begin{aligned}
S_{\mathrm{mag}}^{\mathrm{real}} = \int d\tau d\vec{x}\;&
\Big[ \Pi^i \partial_\tau A_i + \Pi^\tau \partial_\tau A_\tau
- \frac14 F_{ij}F^{ij}\\ &+ A_\tau \partial_i \Pi^i
- u \Pi^\tau \Big] .
\end{aligned}
\tag{3.4} \label{eq:3.4}
\end{equation}
which is precisely the canonical action of magnetic Carroll electrodynamics. All mass terms have dropped out, even though we never set $m=0$ because the scaling exponents automatically suppress them.

Varying action \eqref{eq:3.4} yields several equations of motion:
\begin{itemize}

\item
Variation with respect to $\Pi^i$:
\begin{equation}\partial_\tau A_i - \partial_i A_\tau = 0.\tag{3.5} \label{eq:3.5}
\end{equation}

\item
Variation with respect to $A_i$:
\begin{equation}-\partial_\tau \Pi^{i} + \partial_j F_{ji} = 0.\tag{3.6} \label{eq:3.6}
\end{equation}

\item
Variation with respect to $A_\tau$:
\begin{equation}-\partial_\tau \Pi^{\tau} + \partial_i \Pi^{i} = 0.\tag{3.7} \label{eq:3.7}
\end{equation}

\item
Variation with respect to $\Pi^\tau$:
\begin{equation}\partial_\tau A_\tau = u.\tag{3.8} \label{eq:3.8}
\end{equation}

\item
Variation with respect to $u$:
\begin{equation}\Pi^\tau = 0.\tag{3.9} \label{eq:3.9}
\end{equation}

\end{itemize}
Eq. \eqref{eq:3.9} is the primary constraint. Inserting it into \eqref{eq:3.7} immediately yields the secondary (Gauss) constraint:
\begin{equation}\mathcal{G}_{\mathrm{mag}}=\partial_i \Pi^{i} = 0 .\tag{3.10} \label{eq:3.10}
\end{equation}
Note that, at the extremum of the action, we adopt the notation “=”.
The system is closed by the Bianchi identity for $F_{ij}$. We define the magnetic field $B_k$ and an auxiliary electric vector $E^i$ as
\begin{equation}B_k \equiv \frac12 \epsilon_{kij} F_{ij} = (\nabla \times \vec{A})_k , \qquad
E^i \equiv \Pi^{i} .\tag{3.11} \label{eq:3.11}
\end{equation}
Then the Maxwell equations are
\begin{equation}\begin{aligned}
\nabla \cdot \vec{E} &= 0 ,\;\
\nabla \times \vec{B} + \partial_\tau \vec{E} = 0 ,\\
\nabla \cdot \vec{B} &= 0 ,\;\
\partial_\tau \vec{B} = 0 .
\end{aligned}\tag{3.12} \label{eq:3.12}
\end{equation}

In this magnetic theory, the first‑class constraint is
\begin{equation}\begin{aligned}
\Pi^\tau\approx0,\quad
\mathcal{G}_{\mathrm{mag}} \equiv \partial_i\Pi^{ i} \approx 0.
\end{aligned}
\tag{3.13} \label{eq:3.13}
\end{equation}
generating the U(1) gauge transformations $\delta A_i = \partial_i\epsilon$, $\delta A_\tau=\partial_\tau \epsilon$.

\subsubsection{Contraction induces decoupling with the scalar sector}
It should be noted that in \eqref{eq:3.4}, $A_-$ is no longer present, which of course stems from our choice of scaling exponents. If one wishes to avoid introducing $m$, a viable choice is 
\begin{equation}\begin{aligned}
  \gamma_A &= -1, & \gamma_\Pi &= 0, & \alpha_A &= -2, \\
  \alpha_\Pi &= 1, & \beta_A &= -1, & \beta_\Pi &= 0, \qquad
  \mu = -3.
\end{aligned}
\tag{3.14} \label{eq:3.14}
\end{equation}
And the action reads
\begin{equation}
\begin{aligned}
S_{\mathrm{mag}}^{(1)} = \int d\tau \, d\vec{x} \;
\bigg[ &\; {\Pi}^{i}\partial_\tau A_i  + {\Pi}^{\tau}\partial_\tau A_\tau  - \frac{1}{4}  F_{ij}F^{ij} \\
&+ A_\tau \partial_i {\Pi}^{i} - u {\Pi}^{\tau}\\
&+ {\Pi}^{-}\partial_\tau A_-  - \frac{1}{2}  (\partial_i A_-)(\partial^i A_-) \bigg].
\end{aligned}
\tag{3.15} \label{eq:3.15}
\end{equation}

There are two new equations of motion for the scalar sector:
\begin{equation}
\begin{aligned}
\delta\Pi^{-}&: \quad \partial_{\tau}A_{-}=0,\\
\delta A_{-}&: \quad -\partial_{\tau}\Pi^{-}+\partial_{i}^{2}A_{-}=0 .
\end{aligned}
\tag{3.16} \label{eq:3.16}
\end{equation}
{This is the standard magnetic Carroll scalar, which in principle can evolve self‑consistently. }{However, if one merely wishes to obtain the standard magnetic Carrollian Maxwell theory \eqref{eq:3.4}, one still needs to artificially eliminate this scalar sector—for example, by imposing additional constraints (selecting a suitable solution space) so that the scalar ultimately reduces to zero. Anyway, we would like to mention that this additional constraint is not necessary.} 
A reasonable stable configuration is $\partial_i^2 A_- = 0$ and $\partial_\tau \Pi^- = 0$. {Imposing $\partial_i^2A_-=0$ freezes the dynamics, and by the fall-off condition stated at the beginning of this section the only solution is $A_-=0$. The pair $(A_-, \Pi^-)$ carries no constraint of its own, so any restriction on it is put in by hand. For Eq. \eqref{eq:3.15} the canonical Hamiltonian is $H=\int d\vec{x}\,[\frac14F_{ij}F^{ij}-A_\tau\partial_i\Pi^i+\frac12(\partial_iA_-)(\partial^iA_-)]$. Both $\partial_i^2 A_- \approx 0$ and $A_- \approx 0$ are first class here. They are built from $A_-$, and $\{A_-,H\}=0$ because this $H$ contains no $\Pi^-$. They also commute with $\Pi^\tau$ and $\mathcal{G}$, so the chain terminates at the constraint that is imposed. Other constraints are available as well, but $\Pi^- \approx 0$ is not first class, since $\{\Pi^-(\vec{x}), H\} = \partial_i^2A_-(\vec{x})$. A restriction of that type would instead make the pair second class. We shall not dwell on this, since within the coordinate type the specific choice is free.}
Thus $A_{-}$ describes a frozen, non‑dynamical scalar field. 

What's more, {irrespective of whether the solution space of $A_-$ is artificially selected, }the gauge fields $A_{i},\Pi^{i}$ and the Gauss constraint are completely decoupled from the scalar sector. {Two symmetries act on the reduced theory. The original $U(1)$ gauge symmetry inherited from the KK-like null reduction of the parent theory (here called the original $U(1)$ gauge symmetry) acts on $A_i$ and $A_\tau$ only, and it leaves $A_-$ invariant. The additional symmetry generated by the artificially chosen constraint (here called the additional gauge symmetry) depends on which constraint is used. For $\partial_i^2A_-\approx0$, see Appendix \ref{app:a}, it acts on $\Pi^-$ and again leaves $A_-$ invariant. For the electric scalar of Eq. \eqref{eq:3.27}, where $\Pi^- \approx 0$ is first class, it acts on $A_-$ itself, so $A_-$ is then a pure gauge mode rather than a gauge-invariant field.}

In standard magnetic Carroll Maxwell theory the phase space consists of $2d$ real fields ($A_i, A_\tau, \Pi^i, \Pi^\tau$) and contains two first-class constraints, leaving $(d-2)$ physical degrees of freedom per spatial point. Adding the decoupled scalar field $A_-$ enlarges the phase space to $2(d+1)$ dimensions while the number of first-class constraints stays two, so the physical degrees of freedom become $(d-1)$, the extra one corresponding to a Carrollian scalar mode. However, in Appendix \ref{app:a}, we give the canonical action after {unnecessarily additional} constraint applied to Eq. \eqref{eq:3.15}, where this scalar mode becomes frozen and the physical degrees of freedom reduces to $(d-2)$. {In Section \ref{sec:4}, through a comparison with Ref. \cite{Majumdar:2025juk}, we will discuss the specific meaning of the emergence of the scalar field $A_-$.}
 
{Finally, we emphasize that, without explicitly setting $m=0$, one can obtain a decoupled action in which a magnetic Carroll scalar field coexists with a magnetic Carroll gauge field.}

\subsubsection{No Contraction induces coupling with the scalar sector}

In the magnetic Carrollian theory, the survival of the vector kinetic term $\Pi'^i \partial_\tau A'_i$ and the magnetic energy forces the scaling exponents $\gamma_A = -1$ and $\gamma_\Pi = 0$, completely fixing the scaling of the gauge sector.
The {only mixing term between the vector momentum and the scalar without mass parameter} is $c^{2+\gamma_\Pi+\beta_A}\,\operatorname{Re}\!\big(\Pi'^{\,i}\partial_i A'_-\big)$. With $\gamma_\Pi=0$, a finite contribution in the limit requires its exponent to vanish, which demands $\beta_A = -2$.

However, the scalar gradient energy $|\partial_i A'_-|^2$ carries the exponent $2+2\beta_A$. Setting $\beta_A = -2$ makes this term diverge as $c^{-2}$ when $c\to0$, rendering the action ill‑defined. Regularity of the action then forces $\partial_i A'_- = 0$, which eliminates the very coupling term one tried to preserve. {In fact, the presence of mixing terms does not guarantee genuine coupling, as we shall see shortly in the pseudo-coupled theory obtained from the electric contraction.}

Consequently, within the magnetic Carrollian framework, no finite interacting theory with a dynamical coupling between the magnetic photon and a scalar can be achieved through a consistent $c\to0$ scaling limit.

\subsection{Electric contraction}

\subsubsection{Standard contraction}
To preserve the electric energy $\propto\Pi^{i}\Pi_i$ while discarding the magnetic energy, \begin{equation}\begin{aligned}
  \gamma_A &= 0, & \gamma_\Pi &= -1, & \alpha_A &= -1, \\
  \alpha_\Pi &= 0, & \beta_A &= 0, & \beta_\Pi &= 0, \qquad
  \mu = -2.
\end{aligned}
\tag{3.17} \label{eq:3.17}
\end{equation}
The $c\to0$ limit yields the real action:
\begin{equation}\begin{aligned}
S_{\mathrm{elec}} = \int d\tau d\vec{x}\;&
\Bigl[ \!\left(\Pi^{i} \partial_\tau A_i\right) + \!\left(\Pi^{\tau} \partial_\tau A_\tau\right)\\& - \frac{1}{2} \Pi^{i}\Pi_{i} + A_\tau \partial_i \Pi^{i}  - u \Pi^{\tau}   \Bigr].
\end{aligned}
\tag{3.18} \label{eq:3.18}
\end{equation}

Varying action \eqref{eq:3.18} yields several equations of motion:
\begin{itemize}

\item
Variation with respect to $\Pi^i$:
\begin{equation}
\partial_\tau A_i - \partial_i A_\tau - \Pi^{i} = 0 .
\tag{3.19} \label{eq:3.19}
\end{equation}

\item
Variation with respect to $A_i$:
\begin{equation}
\partial_\tau \Pi^{i} = 0 .
\tag{3.20} \label{eq:3.20}
\end{equation}

\item
Variation with respect to $A_\tau$:
\begin{equation}
-\partial_\tau \Pi^{\tau} + \partial_i \Pi^{i} = 0 .
\tag{3.21} \label{eq:3.21}
\end{equation}

\item
Variation with respect to $\Pi^{\tau}$:
\begin{equation}
\partial_\tau A_\tau - u = 0 .
\tag{3.22} \label{eq:3.22}
\end{equation}

\item
Variation with respect to $u$:
\begin{equation}
\Pi^{\tau}= 0 .
\tag{3.23} \label{eq:3.23}
\end{equation}

\end{itemize}

If we define $E_i \equiv \Pi^i,\ B_k \equiv \frac12 \epsilon_{kij} F_{ij}$, then we can get 
\begin{equation}
\begin{aligned}
\nabla \cdot \vec{E} &= 0 ,\qquad\;\ \
\partial_\tau \vec{E} = 0 ,\\
\nabla \times \vec{E} &= \partial_\tau \vec{B} ,\quad
\nabla \cdot \vec{B} = 0. \quad (\text{Bianchi})
\end{aligned}
\tag{3.24} \label{eq:3.24}
\end{equation}

In this electric theory, the first‑class constraints are still
\begin{equation}\begin{aligned}
\Pi^\tau\approx0,\quad
\mathcal{G}_{\mathrm{elec}} \equiv \partial_i\Pi^{ i} \approx 0.
\end{aligned}
\tag{3.25} \label{eq:3.25}
\end{equation}
which generate the U(1) gauge transformations $\delta A_i = \partial_i\epsilon$, $\delta A_\tau=\partial_\tau \epsilon$.

\subsubsection{Contraction induces decoupling with the scalar sector}

If we choose the following scaling:
\begin{equation}\begin{aligned}
  \gamma_A &= 0, & \gamma_\Pi &= -1, & \alpha_A &= -1, \\
  \alpha_\Pi &= 0, & \beta_A &= 0, & \beta_\Pi &= -1, \qquad
  \mu = -2,
\end{aligned}
\tag{3.26} \label{eq:3.26}
\end{equation}
we can preserve {the standard electric Carrollian scalar $A_-$} in a decoupled way:
\begin{equation}\begin{aligned}
S_{\mathrm{elec}}^{(1)} = \int d\tau d\vec{x}\;
\Big[\, \Pi^{i}\partial_\tau A_i + \Pi^{\tau}\partial_\tau A_\tau 
+ \Pi^{-}\partial_\tau A_- \\
- \frac12 \Pi^{i}\Pi_{i} -\frac{1}{2}\Pi^- \Pi^-
+ A_\tau \partial_i \Pi^{i} - u \Pi^{\tau} \Big] .
\end{aligned}
\tag{3.27} \label{eq:3.27}
\end{equation}

The equations of motion are
\begin{equation}\begin{aligned}\partial_\tau A_- = \Pi^-,\quad
\partial_\tau \Pi^- = 0 \quad\Rightarrow\quad \partial_\tau^2 A_- = 0 .\end{aligned}
\tag{3.28} \label{eq:3.28}
\end{equation}
{For the electric scalar sector a similar restriction is available. The suitable additional constraint is $\Pi^- \approx 0$, which is first class. For Eq. \eqref{eq:3.27} the canonical Hamiltonian is $H=\int d\vec{x}\,[\frac12\Pi^{i}\Pi_{i}+\frac12(\Pi^-)^2-A_\tau\partial_i\Pi^i]$, and $\Pi^-$ has a vanishing Poisson bracket with itself, so $\{\Pi^-(\vec{x}), H\} = 0$. The constraint generates local shifts of $A_-$ through $\delta A_-(\vec{x}) = \{A_-(\vec{x}), \int d\vec{y}\,\epsilon(\vec{y})\,\Pi^-(\vec{y})\} = \epsilon(\vec{x})$, so after gauge reduction the scalar pair $(A_-, \Pi^-)$ is removed and the count of degrees of freedom agrees with the magnetic case. Note also that the constraint $\partial_i^2A_-\approx0$, which works for the magnetic scalar, fails here. Its bracket with $H$ gives $\{\partial_i^2A_-(\vec{x}), H\} = +\partial_i^2\Pi^-(\vec{x})$, so the further constraint $\partial_i^2\Pi^-\approx0$ would be needed, and the two would form a second-class pair. If one imposes $A_- \approx 0$ instead, the set $\{A_- \approx 0,\; \Pi^- \approx 0\}$ is again second class and generates no additional gauge symmetry. Here $\{A_-(\vec{x}),\Pi^-(\vec{y})\}=\delta(\vec{x}-\vec{y})$ does not vanish, while the bracket of each constraint with $H$ vanishes on the constrained surface, so the chain terminates. We present the procedure of imposing ad hoc constraints on the electric scalar theory in Appendix \ref{app:a}.}  

If we choose the following scaling:
\begin{equation}\begin{aligned}
  \gamma_A &= 0, & \gamma_\Pi &= -1, & \alpha_A &= -1, \\
  \alpha_\Pi &= 0, & \beta_A &= -1, & \beta_\Pi &= 0, \qquad
  \mu = -2,
\end{aligned}
\tag{3.29} \label{eq:3.29}
\end{equation}
we can also obtain another action preserving the magnetic scalar sector $A_-$ in a decoupled way:
\begin{equation}
\begin{aligned}
S_{\mathrm{elec}}^{(2)} = \int d\tau d\vec{x} \,
\Big[\, 
&\Pi^{i}\partial_\tau A_i + \Pi^{\tau}\partial_\tau A_\tau
+ \Pi^{-}\partial_\tau A_- 
 \\
-&\frac{1}{2}\Pi^{i}\Pi_{i}
- \frac{1}{2} (\partial_i A_-)(\partial^i A_-) \\
-& \Pi^{i}\partial_i A_- 
+\, A_\tau \partial_i\Pi^{i} 
- u \Pi^{\tau} \Big] .
\end{aligned}
\tag{3.30} \label{eq:3.30}
\end{equation}
What's more, $A_i$ and $A_\tau$ still share the {original} U(1) gauge transformation, while $A_-$ is gauge invariant under this original $U(1)$ transformation in Eq. \eqref{eq:3.27} and \eqref{eq:3.30}. We give a canonical action after {the same additional}
constraint $\partial^2_i A_-=0$ applied to Eq. \eqref{eq:3.30} in Appendix \ref{app:a}. Clearly, before imposing the {additional} constraints, the degrees of freedom in both cases are $(d-1)$; after the additional constraints are imposed, the physical degrees of freedom reduces to $(d-2)$.

\subsubsection{Contraction induces {pseudo-coupling} with the scalar sector}

We choose the following scaling: 
\begin{equation}\begin{aligned}
\gamma_A &= 0, & \gamma_\Pi &= -1, & \alpha_A &= 0, \\
  \alpha_\Pi &= 0, & \beta_A &= -1, & \beta_\Pi &= 0, \qquad
  \mu = 0,
\end{aligned}
\tag{3.31} \label{eq:3.31}
\end{equation}
to obtain the action preserving scalar sector $A_-$ but in a {pseudo-coupled} way:
\begin{equation}\begin{aligned}
S_{\mathrm{elec}}^{(3)} = \int d\tau d\vec{x} \;
\Big[\, \Pi^{i}\partial_\tau A_i + \Pi^{-}\partial_\tau A_- 
- \frac{1}{2} \Pi^{i}\Pi_{i} \\- \frac{1}{2} (\partial_i A_-)(\partial^i A_-) 
- \Pi^{i}\partial_i A_- \Big] .
\end{aligned}
\tag{3.32} \label{eq:3.32}
\end{equation}
The equations of motion are
\begin{equation}\begin{aligned}
\delta\Pi^i &: \;\partial_\tau A_i = \Pi^{i} + \partial_i A_- ,\\
\delta A_i   &: \;-\partial_\tau \Pi^{i} = 0 ,\\
\delta\Pi^- &: \;\partial_\tau A_- = 0 ,\\
\delta A_-   &: \;-\partial_\tau \Pi^- + \partial_i^2 A_- + \partial_i \Pi^{i} = 0 .
\end{aligned}
\tag{3.33} \label{eq:3.33}
\end{equation}

{We note that the apparent coupling between the $A_-$ and $A_i$ in Eq. \eqref{eq:3.32} can be removed by a canonical transformation. Introducing $P^i = \Pi^i + \partial^i A_-$ and $P^- = \Pi^- - \partial_i A^i$, the action reduces to}
\begin{equation}\begin{aligned}
S^{(4)}_{\text{elec}} = \int d\tau d\vec{x}\Bigl[ P^i\partial_\tau A_i + P^-\partial_\tau A_- - \frac12 P^i P_i \Bigr],\end{aligned}
\tag{3.34} \label{eq:3.34}
\end{equation}
{One can verify that the canonical 1‑form
$$\theta = \int d\vec{x}\,\bigl(\Pi^i\,\delta A_i + \Pi^-\,\delta A_-\bigr)$$
becomes, after the substitution,
$$\theta = \int d\vec{x}\,\bigl(P^i\,\delta A_i + P^-\,\delta A_-\bigr) + \delta\!\int d\vec{x}\,A_-\,\partial_i A^i,$$
where the second term is a total variation and therefore does not contribute to the symplectic form $\Omega = \delta\theta$. Hence, the transformation is strictly canonical.}

{And the equations of motions read}
\begin{equation}\begin{aligned}
\delta P^i &: \;\partial_\tau A_i = P_i ,\\
\delta A_i &: \;\partial_\tau P^i = 0 ,\\
\delta P^- &: \;\partial_\tau A_- = 0 ,\\
\delta A_- &: \;\partial_\tau P^- = 0 .
\end{aligned}
\tag{3.35} \label{eq:3.35}
\end{equation}
{This implies that the theory is completely free. And this is the reason why we refer to the original theory in Eq. \eqref{eq:3.32} as “pseudo‑coupling”. Besides, after the canonical transformation, from the action~\eqref{eq:3.34} and the equations of motion~\eqref{eq:3.35}, one may either regard the index of $A_i$ as a spatial index of a vector field, thereby identifying the global $SO(d-1)$ index rotation invariance as the $SO(d-1)$ rotational symmetry of the vector components, or treat the index as an internal index labeling a set of scalar fields, in which case the global $SO(d-1)$ index rotation invariance is recognized as an internal $SO(d-1)$ symmetry. For our purposes, we choose to treat $A_i$ as a spatial vector. Therefore, this symmetry contributes to the total angular momentum of the space vector as the spin part. What's more, the $(A_-, \Pi^-)$ canonical pair becomes a new scalar pair that contributes no energy and is neither a standard electric scalar nor a standard magnetic scalar. In Appendix \ref{app:b}, we verify the theory in Eq. \eqref{eq:3.34} is a Carroll theory.}

It should be emphasized that the theory in \eqref{eq:3.32} or \eqref{eq:3.34} seems no longer to possess any U(1) gauge symmetry. The phase space has $2d$ dimensions and carries no constraint, leaving $d$ degrees of freedom in total. {However, similar to our earlier treatment of the magnetic scalar theory, we may choose to impose additional ad hoc constraints in an attempt to enhance the symmetry of the theory. From $\partial_\tau A_- = 0$ and $\partial_\tau \Pi^i = 0$, $\partial_\tau \Pi^-$ becomes a purely spatial function. So one possible choice is}
\begin{equation}\begin{aligned}\mathcal{G} \equiv \partial_i \Pi^i + \partial_i^2 A_- \approx 0,\end{aligned}
\tag{3.36} \label{eq:3.36}
\end{equation}
{This constraint enforces $\partial_\tau \Pi^- = 0$, which corresponds to a natural restriction on the solution space.}
It generates the additional gauge symmetry acting on the gauge fields as $\delta A_i = \partial_i \epsilon$ while leaving $A_-$ unchanged. We normalize the parameter $\epsilon$ so that the transformation of $A_i$ carries the same sign as the original $U(1)$ transformations of Eqs. \eqref{eq:3.13} and \eqref{eq:3.25}. 

One can rearrange the action by introducing new multipliers $\lambda$ and $v$:
\begin{equation}
\begin{aligned}
S_{\text{elec}}^{(3')} = \int d\tau d\vec{x} \;
\Big[\, 
&\Pi^{i}\partial_\tau A_i + \Pi^{-}\partial_\tau A_- + \Pi^\lambda \partial_\tau \lambda \\
-&\frac12 \Pi^{i}\Pi_{i} - \frac12 (\partial_i A_-)(\partial^i A_-) - \Pi^{i}\partial_i A_- \\
-&\lambda\,\bigl(\partial_i \Pi^{i} + \partial_i^2 A_-\bigr) - v\,\Pi^\lambda \Big] .
\end{aligned}\tag{3.37} \label{eq:3.37}\end{equation}
And the generator for the gauge transformation is
\begin{equation}
\begin{aligned}
G[\epsilon] = \int d\vec{x} \, \big[ -\epsilon\,\mathcal{G} - (\partial_\tau \epsilon\,)\Pi^\lambda \big],\end{aligned}\tag{3.38} \label{eq:3.38}\end{equation}
with
\begin{equation}
\begin{aligned}
\delta A_i &= \partial_i \epsilon , \quad & \delta A_- &= 0 , \quad & \delta \lambda &= -\partial_\tau \epsilon ,\\
\delta \Pi^i &= 0 , \quad & \delta \Pi^- &= \nabla^2 \epsilon , \quad & \delta \Pi^\lambda &= 0 ,\\
\delta v &= -\partial_\tau^2 \epsilon . & &
\end{aligned}\tag{3.39} \label{eq:3.39}\end{equation}

The overall sign of the gauge parameter is fixed throughout by requiring $\delta A_i=\partial_i\epsilon$, as in Eqs. \eqref{eq:3.13} and \eqref{eq:3.25}. The remaining signs in Eq. \eqref{eq:3.39} then follow from the invariance of Eq. \eqref{eq:3.37}.

The phase space consists of $2(d+1)$ real fields per spatial point, and the two first-class constraints $\Pi^\lambda \approx 0$ and $\mathcal{G} \approx 0$ reduce the physical degrees of freedom to $\frac12\big[2(d+1)-2\times 2\big] = (d-1)$. The chain closes at these two, since $\mathcal{G}$ commutes with the Hamiltonian of Eq. \eqref{eq:3.37} and with $\Pi^\lambda$. 

{After the canonical transformation, the additional constraint \eqref{eq:3.36} becomes equivalent to the Gauss constraint $\partial_i P^i = 0$. Therefore, when we add an additional constraint, $S^{(4)}_{elec}$ can also be rearranged by introducing new multipliers as follows}
\begin{equation}
\begin{aligned}
S^{(4')}_{\text{elec}} = \int d\tau d\vec{x} \Big[ &P^i \partial_\tau A_i + P^- \partial_\tau A_- + \Pi^\lambda \partial_\tau \lambda \\
&- \frac12 P^i P_i - \lambda (\partial_i P^i) - v \Pi^\lambda \Big].
\end{aligned}\tag{3.40} \label{eq:3.40}\end{equation}
{Indeed, this is precisely the standard electric Carroll Maxwell theory with a Gauss constraint in Eq. \eqref{eq:3.18}, together with a decoupled zero-energy scalar $A_-$ in which the originally active multiplier $A_\tau$ is replaced by $\lambda$. This means that the constraints we imposed artificially now turn out to be essential components of the standard gauge theory.  Therefore, just as in the earlier treatment of the decoupled magnetic scalar field, imposing additional ad hoc constraints can bring the theory back to the standard electric or magnetic Maxwell theory as much as possible. However, if the electric scalar $A_-$ emerges instead, the first-class constraint $\Pi^- \approx 0$ can be used to eliminate the scalar. We emphasize once again that the choice of such ad hoc constraints is not necessary.}

\subsection{Pure scalar contraction}

We can choose the following scalings:
\begin{equation}\begin{aligned}
\gamma_A &= 0, & \gamma_\Pi &= 0, & \alpha_A &= 0, \\
  \alpha_\Pi &= 0, & \beta_A &= -1, & \beta_\Pi &= 0, \qquad
  \mu = 0,
\end{aligned}
\tag{3.41} \label{eq:3.41}
\end{equation}
and obtain {the standard magnetic} scalar theory:
\begin{equation}\begin{aligned}S_{\text{mag scalar}} = \int d\tau d\vec{x} \;
\Big[\, \Pi^{-}\partial_\tau A_- - \frac12\, (\partial_i A_-)(\partial^i A_-) \,\Big].\end{aligned}
\tag{3.42} \label{eq:3.42}
\end{equation}
The equations of motion are exactly Eq. \eqref{eq:3.16}. We give the canonical action after applying {additional} constraint to Eq. \eqref{eq:3.42} in Appendix \ref{app:a}.

{Besides, we can choose the following scalings:}
\begin{equation}\begin{aligned}
\gamma_A &= 1, & \gamma_\Pi &= 1, & \alpha_A &= 0, \\
  \alpha_\Pi &= 1, & \beta_A &= 0, & \beta_\Pi &= -1, \qquad
  \mu = 0,
\end{aligned}
\tag{3.43} \label{eq:3.43}
\end{equation}
{and obtain the standard electric scalar theory} 
\begin{equation}\begin{aligned}S_{\text{elec scalar}} = \int d\tau d\vec{x}\; \Big[ \Pi^- \partial_\tau A_- - \frac12 (\Pi^-)^2 \Big].\end{aligned}
\tag{3.44} \label{eq:3.44}
\end{equation}
{The equations of motion are Eq. \eqref{eq:3.28}. Also here the additional constraint $\Pi^- \approx 0$ is first class, so it can be imposed to enhance the gauge symmetry of the electric scalar.}

We collect the options for the additional constraint on the scalar pair. Every restriction on the scalar pair is optional and is put in by hand. Which constraint is first class depends on the type of the scalar. For the magnetic scalar of Eqs. \eqref{eq:3.15} and \eqref{eq:3.30} the canonical Hamiltonian is $H=\int d\vec{x}\,[\frac14F_{ij}F^{ij}-A_\tau\partial_i\Pi^i+\frac12(\partial_iA_-)(\partial^iA_-)]$ for Eq. \eqref{eq:3.15} and $H=\int d\vec{x}\,[\frac12\Pi^{i}\Pi_{i}-A_\tau\partial_i\Pi^i+\frac12(\partial_iA_-)(\partial^iA_-)+\Pi^{i}\partial_iA_-]$ for Eq. \eqref{eq:3.30}. Both are built from $A_-$ and its derivatives, so $\partial_i^2A_-\approx0$ and $A_- \approx 0$ are first class, while $\Pi^- \approx 0$ is not. Consistency forces a secondary constraint, and the pair formed by $\Pi^-\approx0$ and that secondary constraint is second class because its members do not commute. That pair still removes the scalar mode. For Eq. \eqref{eq:3.15} the secondary is $\partial_i^2A_-\approx0$, whereas for Eq. \eqref{eq:3.30} the extra term $\Pi^i\partial_iA_-$ makes the secondary $\partial_i^2A_- + \partial_i\Pi^i \approx 0$. For the electric scalar of Eq. \eqref{eq:3.27} the Hamiltonian is built from $\Pi^-$ alone, so the roles are reversed. There $\Pi^- \approx 0$ is first class and generates local shifts of $A_-$, while $\partial_i^2A_-\approx0$ needs the secondary $\partial_i^2\Pi^-\approx0$ and the pair is second class. In every case the scalar pair is removed. The counting uses the fall-off condition stated in Section \ref{sec:3}, which excludes the harmonic modes of $A_-$ and $\Pi^-$. In the theories that contain the vector field this leaves $(d-2)$ physical degrees of freedom per spatial point, while in the pure scalar theories of Eqs. \eqref{eq:3.42} and \eqref{eq:3.44} it leaves none.

\begin{figure}[h]
    \centering
\includegraphics[width=0.9\linewidth]{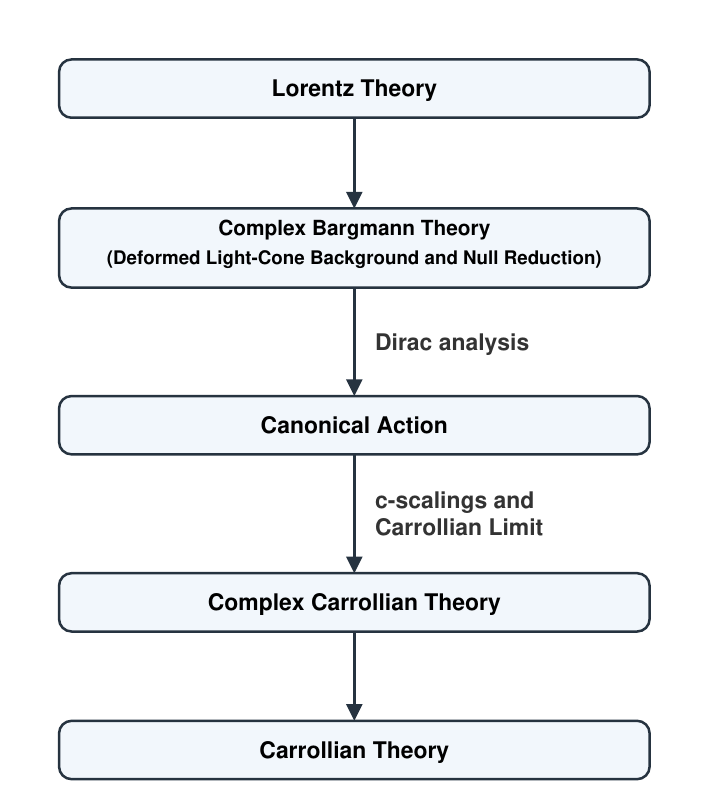}
    \caption{Workflow of deformed light-cone KK-like null reduction.}
    \label{fig:1}
\end{figure}

\section{Discussion}
\label{sec:4}

In this section, we present the procedure for obtaining Carrollian theories via deformed light-cone KK-like null reduction, aiming to clarify potential misconceptions arising from deriving Carrollian theories directly from a pure Lagrangian approach. {What is more, we will discuss the physical implications of the ad hoc constraints employed in the non‑standard magnetic and electric Carroll theories presented in Section \ref{sec:3}.}

Ref. \cite{Saha:2025wkr} outlined a specific procedure for deriving the Carrollian scalar theory from a scalar Lagrangian, in which the scaling of fields relies entirely on the Lagrangian framework. Following the approach of \cite{Saha:2025wkr}, Ref. \cite{Zeng:2026zcj} performed a similar reduction for a complex vector theory. However, surprisingly, the resulting Carrollian theory turned out to be a collection of free complex scalar fields lacking gauge symmetry, with an increased number of physical degrees of freedom due to the loss of symmetry constraints. Based on a framework involving independent scaling of different field components of the complex vector field, Ref. \cite{zeng2026gauge} derived a no-go theorem asserting that it is impossible to obtain a Carrollian theory preserving the gauge structure through mere rescaling of field components. 

In light of these issues, we propose an alternative strategy: first performing the KK-like null reduction, then conducting a Dirac analysis based on the reduced Lagrangian to derive the complete set of constraints, and subsequently constructing the canonical action. We then apply independent scalings to the canonical variables in phase space, thereby obtaining a Carrollian theory with a well-defined structure. Since the introduction of complex fields can be regarded as originating from the complexification of an underlying real theory, we finally convert the resulting complex Carrollian theory back into its real counterpart. Refer to the workflow diagram shown in Fig. \ref{fig:1}. We have already illustrated this for Maxwell theory in Section \ref{sec:2} and \ref{sec:3}. However, it is worth emphasizing that the resulting Carrollian theory depends on the specific choice of scaling.

{With certain choices of scalings we obtain non‑standard Carrollian Maxwell theories with a scalar $A_-$. When $A_-$ appears as a decoupled magnetic scalar, see Eq. \eqref{eq:3.15} and \eqref{eq:3.30}, we introduce an additional ad hoc first‑class constraint and select the solution space in which the dynamics of $A_-$ is frozen; this artificial constraint also brings in an additional gauge symmetry. When $A_-$ appears as a decoupled electric scalar, see Eq. \eqref{eq:3.27}, a similar first-class constraint is available, namely $\Pi^- \approx 0$, and the scalar pair is again removed. The type of the constraint is fixed by the type of the scalar. The magnetic scalar admits constraints built from $A_-$, such as $\partial_i^2A_-\approx0$ or $A_-\approx0$, whereas the electric scalar admits those built from $\Pi^-$. It is worth noting that in the electric contraction $A_-$ can emerge either as a magnetic scalar, Eq. \eqref{eq:3.30}, or as an electric one, Eq. \eqref{eq:3.27}. The pair of Eq. \eqref{eq:3.34} is instead a zero-energy scalar, which contributes no energy and is neither a standard electric scalar nor a standard magnetic scalar. The magnetic contraction, by contrast, only yields a magnetic scalar $A_-$, Eq. \eqref{eq:3.15}. What's more, when $A_-$ is a zero-energy scalar, there is no need to add an additional constraint on $A_-$.
In the pseudo-coupled case of Eq. \eqref{eq:3.32}, adding an additional Gauss constraint on $(A_i,P^i)$ makes the electric vector $A_i$ behave like an electric Maxwell theory.}

\begin{figure}[h]
    \centering
\includegraphics[width=1\linewidth]{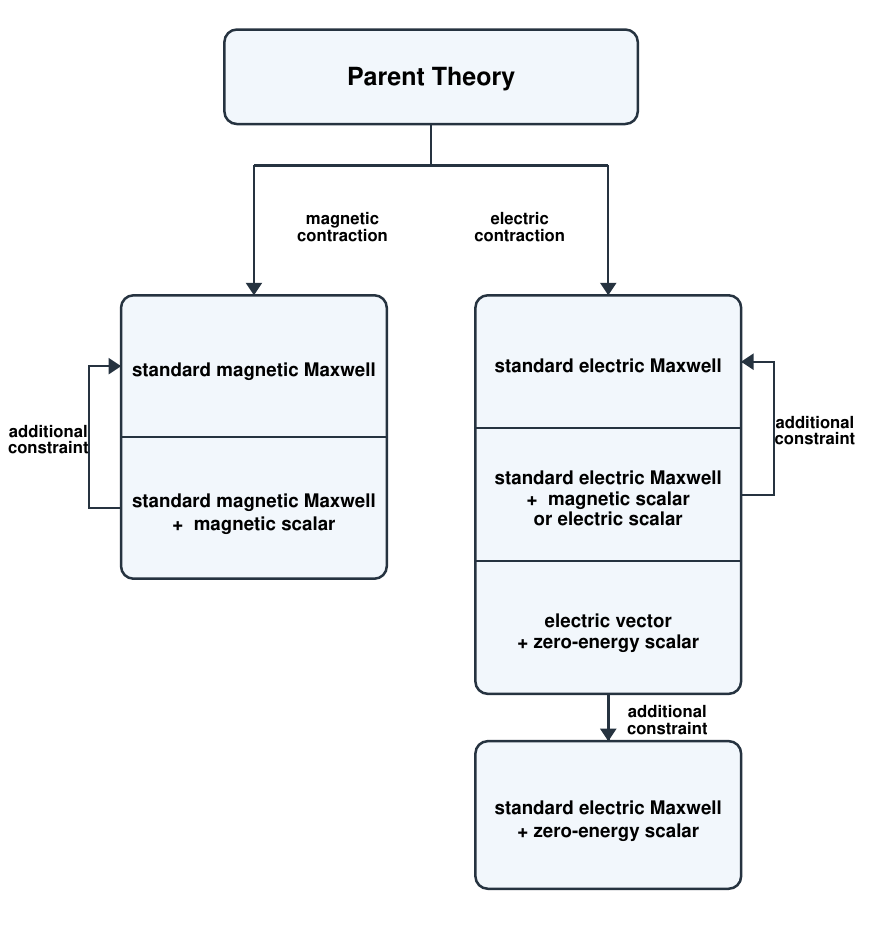}
    \caption{Relations among the non‑trivial theories of Section \ref{sec:3} that contain a vector field. The arrows marked ``additional constraint'' are optional. On the left the constraint is $\partial_i^2A_-\approx0$. On the right there are two, $\Pi^- \approx 0$ and the Gauss constraint on $(A_i,P^i)$.}
    \label{fig:2}
\end{figure}

{To obtain the standard electric and magnetic Carroll theories, Ref. \cite{Majumdar:2025juk} imposes the light‑cone gauge condition $A_- = 0$ while performing the null reduction. In contrast, our approach preserves manifest gauge symmetry by not imposing any additional gauge‑fixing condition on the original $U(1)$ symmetry either before or after the null reduction. As a consequence, a dynamical scalar field $A_-$ naturally appears in our deformed Carrollian theories. One may artificially impose additional constraints on $A_-$, although this is by no means necessary. Furthermore, since different scaling choices are allowed in principle, one can also completely strip off $A_i$, thereby obtaining a pure magnetic or electric Carroll scalar theory consisting solely of the scalar field $A_-$. In Fig. \ref{fig:2} we summarize the relations among the non‑trivial theories of Section \ref{sec:3} that contain a vector field.}

Although our approach effectively preserves the phase space structure, it sacrifices the convenience of dynamically realizing Carrollian conformal symmetry.

\section{Conclusion}
\label{sec:5}
In Section \ref{sec:2}, we first derive the effective action via deformed light-cone KK-like null reduction, and then obtain the canonical action with manifest U(1) gauge symmetry through Dirac analysis. In Section \ref{sec:3}, we begin by introducing a general ansatz that assigns distinct scaling powers of $c$ to each pair of canonical variables, from which we successfully recover both the standard magnetic and electric Carrollian Maxwell theories. However, by adopting alternative admissible scaling exponents for $c$, we obtain modified versions of both the magnetic and electric Carrollian theories that feature an additional scalar sector $A_-$ in their actions. {For the magnetic Carrollian theory, only magnetic scalar $A_-$ would appear in a decoupled manner. For the electric Carrollian theory, the decoupled magnetic or electric scalar $A_-$ both could appear. There also exists an interesting $A_i$–$A_-$ pseudo-coupled electric theory without gauge symmetry. However, after taking canonical transformation and imposing an ad hoc Gauss constraints to $(A_i,P^i)$, this theory reduces to the standard electric Maxwell theory supplemented by a decoupled zero-energy scalar $A_-$. }{For theories containing the magnetic scalar $A_-$, one may impose additional constraints to freeze the dynamics of $A_-$. The resulting theory is the same as the one reached by adopting the light-cone gauge as shown in Ref. \cite{Majumdar:2025juk}, although here $A_-$ is removed by a constraint and not by a gauge choice.} After imposing these {additional} constraints, the magnetic scalar $A_-$ becomes non‑dynamical, and the total number of physical degrees of freedom reduces to match that of the standard Carrollian gauge theory. For the electric scalar, the constraint $\Pi^- \approx 0$ plays the same role, so the scalar pair can be removed there as well and the same reduction of the degrees of freedom follows. Finally, in Section \ref{sec:4}, we outline a general procedure for implementing deformed light-cone null reduction within the Hamiltonian framework, highlight the relations among the different Carrollian theories obtained via the KK-like null reduction and anticipate that this approach can be extended to non-Abelian gauge fields.

The quantization of these constrained Carrollian theories is left for future work. A consistent treatment requires a specification of the temporal boundary conditions and of the prescription for the zero-frequency modes, which we do not give here. Since \cite{Banerjee:2022ocj,Bergshoeff:2023vfd,bagchi2026,majumdar2026} derived Carrollian fermions, a natural question is whether the deformed light-cone KK-like null reduction method can be extended to construct Carrollian fermion. We would like to leave it as future work.

\begin{acknowledgements}
Limin Zeng would like to express his deepest and foremost thanks to his partner Lu Lu. L. Z. is grateful to Cong Ma (BUPT), Ruirui Wu and Anna Tokareva (HIAS, UCAS) for their special support. Finally, L. Z. would like to extend his gratitude to his family for their warm love.
\end{acknowledgements}

\appendix
\section{Canonical action with additional constraint}
\label{app:a}
In this appendix, we present the canonical action of the non-standard magnetic/electric Carrollian theory discussed in Section \ref{sec:3}, after imposing {additional} constraints.

For Eq. \eqref{eq:3.15}, we {artificially} add an {additional} constraint $\partial_i^2 A_- = 0$, and introduce the multipliers $\sigma$ and $w$:
\begin{equation}
\begin{aligned}
S_{\mathrm{mag}}^{(1')} = \int d\tau d\vec{x}\,
\Big[\, 
&\Pi^i \partial_\tau A_i + \Pi^\tau \partial_\tau A_\tau + \Pi^- \partial_\tau A_- \\+ &\Pi^\sigma \partial_\tau \sigma 
- \frac14 F_{ij}F^{ij} + A_\tau \partial_i \Pi^i \\
-& \frac12 (\partial_i A_-)(\partial^i A_-) - \sigma\,\partial_i^2 A_- \\
-& u\,\Pi^\tau - w\,\Pi^\sigma \Big] .
\end{aligned}
\tag{A.1}\label{eq:A.1}
\end{equation}
The four constraints of this theory are $\Pi^\tau\approx0$, $\partial_i\Pi^i\approx0$, $\Pi^\sigma\approx0$ and $\partial_i^2A_-\approx0$. For Eq. \eqref{eq:A.1} the canonical Hamiltonian is $H=\int d\vec{x}\,[\frac14F_{ij}F^{ij}-A_\tau\partial_i\Pi^i+\frac12(\partial_iA_-)(\partial^iA_-)+\sigma\,\partial_i^2A_-]$. The primary constraint $\Pi^\tau\approx0$ gives $\{\Pi^\tau,H\}=\partial_i\Pi^i$, so $\partial_i\Pi^i\approx0$ is secondary, and $\{\partial_i\Pi^i,H\}=0$. The primary constraint $\Pi^\sigma\approx0$ gives $\{\Pi^\sigma,H\}=-\partial_i^2A_-$, so $\partial_i^2A_-\approx0$ is secondary, and $\{\partial_i^2A_-,H\}=0$. Both chains therefore terminate, and the brackets among the four constraints all vanish, so all four are first class. The phase space of the magnetic Carroll theory with a scalar consists of $2(d+2)$ real fields per spatial point, and $4$ first-class constraints reduce the physical degrees of freedom to $(d-2)$. This count uses the fall-off condition of Section \ref{sec:3}, which removes the harmonic modes of $A_-$ and $\Pi^-$. Without it these modes would survive and the count would have to be modified.

{The generator of gauge transformations is as follows:
$$G[\epsilon,\zeta] = \int d\vec{x}\,
\Big[ (\partial_\tau\epsilon)\Pi^\tau - \epsilon(\partial_i\Pi^i) + (\partial_\tau\zeta)\Pi^\sigma + \zeta(\partial_i^2A_-) \Big]$$ with} 
\begin{equation*}
\begin{aligned}
\delta A_i &= \partial_i\epsilon, & \delta A_\tau &= \partial_\tau\epsilon, \\
\delta u &= \partial_\tau^2\epsilon, & \delta A_- &= 0, \\
\delta \sigma &= \partial_\tau\zeta, & \delta w &= \partial_\tau^2\zeta, \\
\delta \Pi^- &= -\partial_i^2\zeta. & &
\end{aligned}
\end{equation*}
{where $\epsilon$ parametrizes the original U(1) gauge transformation and $\zeta$ parametrizes the additional gauge transformation.} The overall sign of $\zeta$ is fixed in the same way. The additional symmetry of Eq. \eqref{eq:3.39} is written with the parameter $\epsilon$, for which $\delta\Pi^-=+\nabla^2\epsilon$, while here it is written with $\zeta$, for which $\delta\Pi^-=-\partial_i^2\zeta$. The two agree when $\zeta=-\epsilon$.

Similarly, for electric theory in Eq. \eqref{eq:3.30} we {could} also add an {additional} constraint $\partial_i^2 A_- \approx 0$, and introduce the multipliers $\sigma$ and $w$:
\begin{equation}
\begin{aligned}
S_{\text{elec}}^{(2')} = \int d\tau d\vec{x}\,
\Big[\, 
&\Pi^{i}\partial_\tau A_i + \Pi^{\tau}\partial_\tau A_\tau + \Pi^{-}\partial_\tau A_- \\+ &\Pi^\sigma \partial_\tau \sigma 
-\frac{1}{2}\Pi^{i}\Pi_{i} \\-& \frac{1}{2} (\partial_i A_-)(\partial^i A_-) - \Pi^{i}\partial_i A_- 
+ A_\tau \partial_i\Pi^{i} \\-& u \Pi^{\tau} - \sigma\,\partial_i^2 A_- - w\,\Pi^\sigma \Big] .
\end{aligned}\tag{A.2}\label{eq:A.2}\end{equation}
The four constraints of this theory are $\Pi^\tau\approx0$, $\partial_i\Pi^i\approx0$, $\Pi^\sigma\approx0$ and $\partial_i^2A_-\approx0$. For Eq. \eqref{eq:A.2} the canonical Hamiltonian is $H=\int d\vec{x}\,[\frac12\Pi^{i}\Pi_{i}+\frac12(\partial_iA_-)(\partial^iA_-)+\Pi^{i}\partial_iA_- -A_\tau\partial_i\Pi^i+\sigma\,\partial_i^2A_-]$. The two chains are the same as for Eq. \eqref{eq:A.1}, because $\{\Pi^\tau,H\}=\partial_i\Pi^i$ and $\{\Pi^\sigma,H\}=-\partial_i^2A_-$ as before, while $\{\partial_i\Pi^i,H\}=0$ and $\{\partial_i^2A_-,H\}=0$ still hold. All four constraints are therefore first class. The phase space of the electric Carroll theory with a scalar consists of $2(d+2)$ real fields per spatial point, and $4$ first-class constraints reduce the physical degrees of freedom to $(d-2)$. The count again uses the fall-off condition of Section \ref{sec:3}.

Also, the {canonical action for pure scalar theory \eqref{eq:3.42} after imposing additional constraint} is
\begin{equation}
\begin{aligned}
S_{\text{mag scalar}}' =& \int d\tau d\vec{x} \;
\Big[\, \Pi^{-}\partial_\tau A_- + \Pi^\sigma \partial_\tau \sigma
\\&- \frac12 (\partial_i A_-)(\partial^i A_-) - \sigma\,\partial_i^2 A_- - w\,\Pi^\sigma \Big] .
\end{aligned}
\tag{A.3}\label{eq:A.3}\end{equation}
For Eq. \eqref{eq:A.3} the canonical Hamiltonian is $H=\int d\vec{x}\,[\frac12(\partial_iA_-)(\partial^iA_-)+\sigma\,\partial_i^2A_-]$. The constraint $\Pi^\sigma\approx0$ gives $\{\Pi^\sigma,H\}=-\partial_i^2A_-$, so $\partial_i^2A_-\approx0$ is secondary, while $\{\partial_i^2A_-,H\}=0$, so the chain terminates and the two constraints are first class. And the physical degrees of freedom are reduced to $0$.

In fact, all first-class constraints can serve as generators of gauge transformations\cite{Henneaux:1992ig}; therefore, upon imposing {additional} constraints, the gauge symmetry is no longer limited to the {original} U(1) gauge symmetry of the original parent theory. As we have shown in Eq. \eqref{eq:3.34} and Eq. \eqref{eq:3.40}, $S_{\mathrm{elec}}^{(4)}$ itself has no gauge symmetry, but after adding {additional} constraints, a new gauge symmetry manifests in $S_{\mathrm{elec}}^{(4')}$. {Moreover, the theory obtained after imposing the ad hoc constraints is precisely the standard electric Maxwell theory together with a decoupled zero-energy scalar, see Eq. \eqref{eq:3.40}.}

{In the end, we want to show that after imposing $A_- \approx 0$ in the pure electric scalar theory \eqref{eq:3.44}, the resulting constraint pairs are second-class. The full canonical action reads}
\begin{equation}
\begin{aligned}
S'_{\text{elec scalar}} = \int d\tau d\vec{x}\;
\Bigl[\,\Pi^-\partial_\tau A_- + \Pi^\sigma \partial_\tau \sigma\\
- \frac12 (\Pi^-)^2 - \sigma A_- - w\,\Pi^\sigma \Bigr].
\end{aligned}
\tag{A.4}\label{eq:A.4}\end{equation}
{Corresponding constraints are 
$\phi_1=\Pi^\sigma$, $\phi_2=A_-$, $\phi_3=\Pi^-$, $\phi_4=\sigma$. They form a single chain. For Eq. \eqref{eq:A.4} the canonical Hamiltonian is $H=\int d\vec{x}\,[\frac12(\Pi^-)^2+\sigma A_-]$. The primary constraint $\Pi^\sigma\approx0$ gives $\{\Pi^\sigma,H\}=-A_-$, so $A_-\approx0$ is secondary. Then $\{A_-,H\}=\Pi^-$ makes $\Pi^-\approx0$ tertiary, and $\{\Pi^-,H\}=-\sigma$ makes $\sigma\approx0$ quaternary. The chain terminates because $\sigma$ commutes with $H$. The Poisson brackets are}
\begin{equation*}
\begin{aligned}
\{\phi_2(\vec{x}), \phi_3(\vec{y})\} = \{A_-(\vec{x}), \Pi^-(\vec{y})\} = \delta(\vec{x}-\vec{y}) \neq 0,\\
\{\phi_1(\vec{x}), \phi_4(\vec{y})\} = \{\Pi^\sigma(\vec{x}), \sigma(\vec{y})\} = -\delta(\vec{x}-\vec{y}) \neq 0 .
\end{aligned}
\end{equation*}
{Therefore, $(\phi_2, \phi_3)$ and $(\phi_1, \phi_4)$ each form a second-class constraint pair, and the bracket matrix is block diagonal with two non-degenerate $2\times2$ blocks. The pair $(\phi_2,\phi_3)$ is the scalar pair $(A_-,\Pi^-)$ itself. The pair $(\phi_1,\phi_4)$ is the auxiliary pair formed by the multiplier $\sigma$, which imposes $A_-\approx0$, together with its conjugate momentum $\Pi^\sigma$. All four variables are removed and no degrees of freedom remain.} For comparison, imposing only $\Pi^- \approx 0$ in Eq. \eqref{eq:3.44} gives a first-class constraint, since $\Pi^-$ has a vanishing Poisson bracket with itself. The pair $(A_-, \Pi^-)$ is then removed by gauge fixing, and the outcome is the same. The second-class pair of $\{A_- \approx 0, \Pi^- \approx 0\}$ is therefore one choice among others.

\section{Verification of the new Carrollian theory}
\label{app:b}

{We compute step by step the Poisson brackets among the Carroll generators for the unconstrained action in Eq. \eqref{eq:3.34}, thereby verifying that the theory satisfies Carroll symmetry.}

{The phase space variables are $(A_i, P^i)$ and $(A_-, P^-)$, with the fundamental brackets $$\begin{aligned}\{A_i(\vec{x}), P^j(\vec{y})\} &= \delta_i^j\,\delta(\vec{x}-\vec{y}),\\
\{A_-(\vec{x}), P^-(\vec{y})\} &= \delta(\vec{x}-\vec{y}),\end{aligned}$$
and all others vanishing. The Carroll generators are defined as
$$\begin{aligned}
\hat{E} &= \frac12\int d\vec{x}\, P^i P_i, \\
\hat{P}_k &= \int d\vec{x}\,\bigl(P^i\partial_k A_i + P^-\partial_k A_-\bigr), \\
\hat{B}_m &= \frac12\int d\vec{x}\,{x}_m\, P^i P_i, \\
\hat{M}_{rs} &= \int d\vec{x}\,\left[{x}_r P^l\partial_s A_l - {x}_s P^l\partial_r A_l 
            + P_r A_s - P_s A_r\right] \\
            &\quad + \int d\vec{x}\,\bigl({x}_r P^-\partial_s A_- - {x}_s P^-\partial_r A_-\bigr).
\end{aligned}$$
}
{The scalar sector has zero energy, and its boost generator therefore vanishes (so any algebraic identity involving the energy or boost generators reduces to the trivial check of whether the bracket vanishes). The generators of the scalar sector commute completely with those of the vector sector. It therefore suffices to verify the algebra separately for the vector and scalar sectors. In the following we verify the algebra explicitly for the vector sector; for the scalar sector, the checks involving the energy and boost generators are trivially satisfied since those generators contain no scalar field contributions.}

\begin{enumerate}
\item { $[\hat{E}, \hat{P}_k] = 0$:
$$\begin{aligned}
\{\hat{E}, \hat{P}_k\} &=\int d\vec{z}\left[ \frac{\delta \hat{E}}{\delta A_i}\frac{\delta \hat{P}_k}{\delta P^i} - \frac{\delta \hat{E}}{\delta P^i}\frac{\delta \hat{P}_k}{\delta A_i} \right] \\
&= \int d\vec{z}\left[ 0 - P_i(\vec{z})\big(-\partial_k P^i(\vec{z})\big) \right] \\
&= \int d\vec{z}\,P_i\,\partial_k P^i = \frac12\!\int d\vec{z}\,\partial_k P^2 = 0.
\end{aligned}$$}

\item {$[\hat{E}, \hat{B}_m] = 0$:
$$\frac{\delta \hat{B}_m}{\delta P^i(\vec{x})} = {x}_m P_i(\vec{x}),\quad \frac{\delta \hat{B}_m}{\delta A_i}=0.$$
Hence
$$\{\hat{E}, \hat{B}_m\} = \int d\vec{z}\Big[ -\frac{\delta \hat{E}}{\delta P^i}\frac{\delta \hat{B}_m}{\delta A_i} \Big] = 0.$$}

\item {$[\hat{P}_k, \hat{P}_l] = 0$:
$$\frac{\delta \hat{P}_k}{\delta A_i} = -\partial_k P^i, \quad\frac{\delta \hat{P}_k}{\delta P^i} = \partial_k A_i.$$Hence
$$\begin{aligned}
\{\hat{P}_k, \hat{P}_l\} &= \int d\vec{z}\left[ \frac{\delta \hat{P}_k}{\delta A_i}\frac{\delta \hat{P}_l}{\delta P^i} - \frac{\delta \hat{P}_k}{\delta P^i}\frac{\delta \hat{P}_l}{\delta A_i} \right] \\
&= \int d\vec{z}\Big[ (-\partial_k P^i)(\partial_l A_i) - (\partial_k A_i)(-\partial_l P^i) \Big] \\
&=\int d\vec{z}\Big[
\partial_k (A_i \partial_l P^i) - \partial_l (A_i \partial_k P^i)
\Big]=0.
\end{aligned}$$}

\item { $[\hat{P}_k, \hat{B}_m] = \delta_{km}\hat{E}$:
$$\begin{aligned}
\{\hat{P}_k, \hat{B}_m\} &= \int d\vec{z}\left[ \frac{\delta \hat{P}_k}{\delta A_i}\frac{\delta \hat{B}_m}{\delta P^i} - \frac{\delta \hat{P}_k}{\delta P^i}\frac{{\delta \hat{B}_m}}{\delta A_i} \right] \\
&= \int d\vec{z}\,(-\partial_k P^i) (z_m P_i) \\
&= -\frac12\int d\vec{z}\,z_m\,\partial_k(P_i P^i) \\
&= \delta_{km}\,\frac12\int d\vec{z}\,P^2 = \delta_{km}\,E.
\end{aligned}$$
}

\item {$[\hat{M}_{rs}, \hat{P}_k]=\delta_{ks}\hat{P}_r - \delta_{kr}\hat{P}_s$:$$\begin{aligned}
\frac{\delta \hat{M}_{rs}}{\delta A_i(\vec{x})} &= x_s\partial_r {P}^i - x_r\partial_s {P}^i+\delta_s^i P_r(\vec{x}) - \delta_r^i P_s(\vec{x}),\\
\frac{\delta \hat{M}_{rs}}{\delta P^i(\vec{x})} &= x_r\partial_s A_i - x_s\partial_r A_i+\delta_{ri} A_s(\vec{x}) - \delta_{si} A_r(\vec{x}).
\end{aligned}$$
Hence 
$$\begin{aligned}
&\{\hat{M}_{rs}, \hat{P}_k\} =\int d\vec{z}\left[
\frac{\delta \hat{M}_{rs}}{\delta A_i}\frac{\delta \hat{P}_k}{\delta P^i}
- \frac{\delta \hat{M}_{rs}}{\delta P^i}\frac{\delta \hat{P}_k}{\delta A_i}
\right]\\&=\int d\vec{z} \Big[ (z_s \partial_r P^i - z_r \partial_s P^i + \delta_s^i P_r - \delta_r^i P_s) (\partial_k A_i) \\
&\quad- (z_r \partial_s A_i - z_s \partial_r A_i + \delta_{ri} A_s - \delta_{si} A_r) (-\partial_k P^i) \Big] \\
&= \delta_{ks} \int d\vec{z}\, P^i \partial_r A_i - \delta_{rs} \int d\vec{z} \,P^i \partial_k A_i 
  \\&\quad- \bigl( \delta_{kr} \int d\vec{z}\, P^i \partial_s A_i - \delta_{sr} \int d\vec{z} \,P^i \partial_k A_i \bigr) \\
&= \delta_{ks}\hat{P}_r - \delta_{kr}\hat{P}_s.
\end{aligned}$$
}

\item {$[\hat{M}_{rs}, \hat{B}_m]$:
$$\begin{aligned} &\{\hat{M}_{rs}, \hat{B}_m\} = \int d\vec{z}\left[
\frac{\delta \hat{M}_{rs}}{\delta A_i}\frac{\delta \hat{B}_m}{\delta P^i}
- {\frac{\delta \hat{M}_{rs}}{\delta P^i}\frac{\delta \hat{B}_m}{\delta A_i}}
\right]\\
&= \int d\vec{z}\left[ \bigl( z_s \partial_r P^i - z_r \partial_s P^i \bigr) z_m P_i 
   + \bigl( \delta_s^i P_r - \delta_r^i P_s \bigr) z_m P_i\right] \\&= -\delta_{rm} B_s - \delta_{rs} B_m + \delta_{sm} B_r + \delta_{sr} B_m \\
&= \delta_{sm} B_r - \delta_{rm} B_s.\end{aligned}$$}

\end{enumerate}

{Our verification of the Carroll algebra ends here, since the remaining algebraic relations involving angular momentum are standard and can be found in any textbook treatment of the energy‑momentum tensor and angular momentum.}

\clearpage
\bibliography{apssamp}

\end{document}